\documentclass[12pt]{article}

\usepackage{amssymb, amsmath, amsthm}
\usepackage{graphicx}
\usepackage{cite}

\newcommand{\td}{\text{d}}

\usepackage{hyperref}

\def\be{\begin{equation}}
\def\ee{\end{equation}}
\def\bea{\begin{eqnarray}}
\def\eea{\end{eqnarray}}

\theoremstyle{definition}

\usepackage[left=2cm,right=2cm,top=2cm,bottom=2cm]{geometry}

\title{\bf Supersymmetric multi-charge solitons in AdS$_5$}
\author{Turkuler Durgut$^a$\footnote{tdurgut@mun.ca } \  and Hari K. Kunduri$^b$\footnote{hkkunduri@mun.ca } \\ \\
\small \sl $^a$Faculty of Science, Theoretical Physics, Memorial University of Newfoundland \\ \small \sl  St. John's, NL  A1C 5S7, Canada  \\
\small \sl $^b$Department of Mathematics and Statistics, Memorial University of Newfoundland \\ \small \sl 
St. John's, NL  A1C 5S7, Canada }

\date{}

\begin{document}
\maketitle
\begin{abstract} We construct supersymmetric,  asymptotically AdS$_5$ gravitational soliton solutions of five-dimensional gauged supergravity coupled to arbitrarily many vector multiplets.  These generalize the supersymmetric solitons of $U(1)^3$ gauged supergravity previously obtained by Chong, Cvetic, and Pope as limits of local, non-supersymmetric metrics.  We show that the soliton spacetimes contain evanescent ergosurfaces and argue that, as a result, these solitons should be nonlinearly unstable. 
\end{abstract}

\vspace{.5cm}
\section{Introduction} 
A gravitational soliton is a geodesically complete, globally stationary (and horizon-free) non-trivial solution of the Einstein equations, with prescribed asymptotic geometry. It is a classic theorem of Lichnerowicz \cite{Lich} that asymptotically flat, electrovacuum gravitational solitons cannot exist.  The proof is considerably simpler if one employs the spacetime positive mass theorem~\cite{Gibbons:2011aqq}.  In the pure vacuum case, the result extends to higher dimensions~\cite{Shiromizu:2012hb}, and in Einstein-Maxwell theory, static solitons can be ruled out~\cite{Kunduri:2017htl} and no known stationary examples are known (see, e.g. the review \cite{Bena:2007kg}).  On the other hand, within supergravity theories which have Maxwell fields (or higher $p-$form field strengths), there are many known families of asymptotically flat solitons carrying mass, angular momenta, and electric charge `magnetic fluxes' which support non-trivial cycles in the spacetime.  These charges satisfy variational laws analogous to the first law of black hole mechanics~\cite{Kunduri:2013vka}.

The majority of these known solutions are asymptotically flat and supersymmetric (i.e., they are supergravity solutions admitting one or more Killing spinors \cite{Gauntlett:2002nw}). However, this is likely a result of the fact that the integrability conditions imposed by supersymmetry allow for the construction of explicit solutions (see \cite{Breunholder:2017ubu} for classification of supersymmetric black holes and solitons with $U(1)^2$ isometry in minimal ungauged supergravity).  Fewer examples are known in the asymptotically globally AdS setting with conformal boundary $\mathbb{R} \times S^n$.   Non-supersymmetric examples were constructed in \cite{Ross:2005yj} by taking limits of local solutions of gauged supergravity that were first used to obtain charged, rotating asymptotically AdS$_5$ black holes.  As the solitons do not have horizons, they cannot be interpreted as thermal states from the CFT perspective. It is reasonable to interpret them as pure states with non-zero vacuum expectation values for energy and R-charge.

Given the difficulty in constructing solutions of gauged supergravity, it is natural to focus attention on supersymmetric soliton geometries.  A general formalism for constructing BPS solutions of minimal gauged supergravity was given in \cite{Gutowski:2004ez} with the purpose of constructing the first examples of supersymmetric AdS$_5$ black holes.   The analysis shows that BPS solutions can be categorized into either a timelike or null class. Solutions in the timelike class can be constructed by first selecting a four-dimensional K\"ahler base $B$; the remaining field equations are then reduced to geometric equations on $B$.  This work was extended \cite{Gutowski:2004yv} to the more general setting of gauged supergravity coupled to vector multiplets, which contains the $U(1)^3$ gauged supergravity as a particular case. Solutions of the latter theory naturally lift to local solutions of Type IIB supergravity compactified on $S^5$ (or, more generally, a Sasaki-Einstein five-manifold).  This was subsequently used to construct a four-parameter family of $\tfrac{1}{4}-$BPS AdS$_5$ black hole solutions~\cite{Kunduri:2006ek}.   This led to a strict quantitative test of AdS/CFT: could the semi-classical entropy of these black holes be obtained by counting the degeneracy of dual BPS CFT states? Remarkably, this computation was recently achieved through different approaches \cite{Cabo-Bizet:2018ehj,Choi:2018hmj, Benini:2018ywd} (see also the review article \cite{Zaffaroni:2019dhb}).

The above constructions focused mainly on black hole solutions, but the local metrics of \cite{Chong:2005hr} produced as a byproduct the first examples of $\tfrac{1}{2}-$ BPS asymptotically globally AdS$_5$ gravitational solitons.  All the known BPS black holes and solitons have a base space which is a particular member of a class of orthotoric K\"ahler spaces parameterized by two arbitrary functions of a single variable.  A systematic approach to classifying BPS solutions in minimal supergravity with this general class of base spaces was carried out in~\cite{Cassani:2015upa}. This analysis reproduced the known solutions as well as producing new families; in particular, the BPS solitons of \cite{Chong:2005hr}  were rederived and their regularity investigated. The summary was that there is an asymptotically globally AdS$_5$ soliton with positive energy and non-zero angular momenta and electric charge and no free continuous parameters. The spacetime metric possesses an $\mathbb{R} \times SU(2) \times U(1)$ isometry group. Interestingly, the Killing spinors are invariant under a different $SU(2)$ action, and thus if one writes this solution into the canonical supersymmetric form, the base space $B$ does not inherit this symmetry and is merely toric.  Note that recently, a classification of BPS solutions of the minimal theory, which admit an $SU(2)$ action as isometries, was achieved~\cite{Lucietti:2021bbh}. In particular, this work established a uniqueness theorem for the $SU(2)$-invariant BPS AdS$_5$ black holes of \cite{Gutowski:2004ez}.

The purpose of the present work is to construct generalizations of these BPS solitons to the case of gauged supergravity coupled to an arbitrary number of Abelian vector multiplets (referred to as `$U(1)^N$ supergravity').  These new solutions reduce in the case of equal electric charges to the solutions of the minimal theory found by~\cite{Chong:2005hr}.  In the special case of $U(1)^3$ supergravity, our solutions reduce to the ones obtained in \cite{Cvetic:2005zi} by taking a combined BPS and horizonless limit of a local family of $SU(2) \times U(1)$-invariant solutions.  The main difficulty in this construction is that in addition to multiple gauge fields, there are also scalar fields that must be determined.  The supergravity equations couple these fields together, which makes a systematic analysis of all solutions with the given orthotoric base space along the lines of~\cite{Cassani:2015upa} difficult.  We hope to return to a general analysis in the future.

A global analysis of the local metrics reveals a family of everywhere regular solutions parameterized by $N$ positive moduli subject to one constraint. We also demonstrate that they must possess an evanescent ergosurface instability, which to our knowledge is the first such example of this type in the asymptotically AdS setting. This is an instability of stationary solutions associated with the stable trapping of null geodesics near a timelike hypersurface along which the asymptotically stationary Killing field becomes null. A subtlety here is that in asymptotically AdS$_5$ spacetimes, there is more than one such choice of a stationary Killing field, and for a particular choice, this ergosurface is revealed.  A second observation is that the subfamily of solutions of $U(1)^3$-gauged supergravity can indeed be oxidized to globally smooth solutions of Type IIB supergravity on $S^5 / \mathbb{Z}_p, p \geq 3$ provided the moduli parameterizing the solution is suitably quantized (as previously noted in~\cite{Cvetic:2005zi} the case of $S^5$  leads to a global obstruction to smoothness).  The dual CFT duals defined on the $\mathbb{R} \times S^3$ conformal boundary of these geometries are quivered gauge theories.

Our work is organized as follows.  In Section 2, we review the construction of supersymmetric solutions of $U(1)^N$ gauged supergravity and derive our local solutions. We then perform a global analysis of these solutions and compute their conserved charges. We also give a self-contained description of the three-charge BPS soliton solutions of  $U(1)^3-$supergravity and describe their lifting to ten dimensions. Finally, in Section 3, we discuss in some detail the existence of evanescent ergosurfaces and the associated stable trapping on null geodesics. We argue that this provides strong evidence that our solutions must suffer from (at least) a nonlinear instability whose endpoint would be a spacetime containing one or more near-BPS AdS$_5$ black holes.

\label{sec:intro}
\section{Supersymmetric AdS$_5$ solitons}
\subsection{Supersymmetric solutions to gauged supergravity}
The bosonic sector of the theory consists of the metric, $N$ gauge fields $A^I$, and $N-1$ real scalar fields, which are represented by $N$ real scalar fields $X^I$ subject to the constraint~\cite{Gutowski:2004yv, Kunduri:2006ek}
\begin{equation}
\frac{1}{6} C_{IJK} X^I X^J X^K = 1.
\end{equation} 
The $C_{IJK}$ are constants and as a tensor it is totally symmetric, i.e. $C_{IJK} = C_{(IJK)}$ with $I= 1 \ldots N$.  A particular combination that often comes up is 
\begin{equation}
X_I = \frac{1}{6} C_{IJK} X^J X^K
\end{equation} 
The theory is governed by the action~\cite{Kunduri:2006ek} 
\begin{equation}\label{action}
\begin{aligned}
S = \frac{1}{16 \pi G} \int \left(  R \star_5 1 - Q_{IJ} F^I \wedge \star_5 F^J - Q_{IJ} \td X^I \wedge \star_5 \td X^J - \frac{1}{6}  C_{IJK} F^I \wedge F^J \wedge A^K  + 2 g^2 V \star_5 1 \right)
\end{aligned}
\end{equation} 
with $F^I:= \td A^I$ where $A^I$ are local $U(1)$ gauge fields.   The matrix $Q_{IJ}$ is given by
\begin{equation}
Q_{IJ} = \frac{9}{2} X_I X_J - \frac{1}{2} C_{IJK} X^K.
\end{equation} 
The $C_{IJK}$ are assumed to satisfy the following symmetric space condition
\begin{equation}\label{symmcond}
C_{IJK} C_{J'(LM} C_{PQ)K'} \delta^{J J'} \delta^{KK'} = \frac{4}{3} \delta_{I(L} C_{MPQ)}.
\end{equation} 
This condition ensures that $Q_{IJ}$ has an inverse 
\begin{equation}
Q^{IJ} = 2 X^I X^J - 6 C^{IJK} X_K,
\end{equation} 
with the identification $C^{IJK}:= C_{IJK}$.  This also allows us to invert for $X^I$ in terms of the $X_J$:
\begin{equation}
X^I = \frac{9}{2} C^{IJK} X_J X_K,
\end{equation} 
which then implies
\begin{equation}\label{constraint}
C^{IJK} X_I X_J X_K = \frac{2}{9}.
\end{equation} 
Finally the potential 
\begin{equation}
V = 27 C^{IJK} \bar{X}_I \bar{X}_J X_K,
\end{equation} 
where the $\bar{X}_I$ are a set of constants.  As shown in \cite{Gutowski:2004yv},  the vacuum AdS$_5$ background with radius $\ell = 1/g$ corresponds to $A^I\equiv 0$ and constant scalars $X^I = \bar{X}^I$, and  
\begin{equation}
\bar{X}^I \equiv \frac{9}{2} C^{IJK} \bar{X}_J \bar{X}_K.
\end{equation} 
The special $U(1)^3$ supergravity case corresponds to $N=3$, $C_{IJK} =1$ if $(IJK)$ is a permutation of $(123)$ and $C_{IJK}=0$ otherwise and $\bar{X}^I =1$, or equivalently $\bar{X}_I = 1/3$.  The symmetric space condition~\eqref{symmcond} holds automatically. 

Given a Killing spinor, one can show that there is a Killing vector field $V$, which is non-spacelike. So we assume we are in a region where $V^2 = f^2 < 0$ so that $f > 0$ for some function $f$ and the metric can be decomposed as
\begin{equation}\label{BPSmet}
\td s^2 = -f^2 (\td t + \omega)^2 + f^{-1} h_{ab} \td x^a \td x^b,
\end{equation} 
where $V = \partial/ \partial t$. Supersymmetry implies that the 4d metric $h$ is K\"ahler with K\"ahler form $J$, and the orientation of the base space $B$ chosen so that $J$ is anti self dual $\star J = -J$. The 5-form $(\td t + \omega) \wedge \td \text{vol}(h)$ having a positive orientation in the full spacetime. 

The Maxwell field has to take the form
\begin{equation}\label{BPSMaxwell}
F^I = \td \left[ X^I f(\td t + \omega) \right] + \Theta^I - 9 g f^{-1} C^{IJK} \bar{X}_J X_K J,
\end{equation} 
and $\Theta^I$ are self-dual two-forms on $B$, and we must have
\begin{equation}
X_I \Theta^I = -\frac{2}{3} G^+,
\end{equation} 
and $G^\pm$ is the (anti-)self dual two-form with $\star G^\pm = \pm G^\pm$, defined as
\begin{equation}
G^\pm = \frac{1}{2} f ( \td \omega \pm \star \td \omega).
\end{equation} 
Above $\star$ refers to the Hodge dual with respect to $(B, h)$.  This can be inverted so that 
\begin{equation}
\td\omega = f^{-1} (G^+ + G^-).
\end{equation}
Since $(B,h,J)$ is K\"ahler, we can define the Ricci two-form
\begin{equation}
\mathcal{R}_{ab} = \frac{1}{2} R_{ab cd} J^{cd}.
\end{equation} 
Supersymmetry implies that $\mathcal{R} =\td P$ where $P$ is the one-form
\begin{equation}
P = 3 g \bar{X}_I \left( A^I - f X^I \omega\right).
\end{equation} 
This determines completely the function $f$ as 
\begin{equation}\label{BPSf}
f = - \frac{108 g^2}{R} C^{IJK} \bar{X}_I \bar{X}_J X_K,
\end{equation} 
and the following condition holds
\begin{equation}\label{ricciform}
\mathcal{R} - \frac{R}{4} J = 3 g \bar{X}_I \Theta^I.
\end{equation}

All these conditions are necessary and turn out to be sufficient to guarantee and the existence of a supercovariantly constant spinor.  All the field equations are satisfied provided $\td F^I =0$ (which is automatically true if we specify potentials), and the Maxwell equations 
\begin{equation}
\td (Q_{IJ} \star_5 F^J) = -\frac{1}{4} C_{IJK} F^J \wedge F^K
\end{equation} are satisfied.
The Bianchi identity and the Maxwell equation respectively reduce to the following equations on the base space
\begin{equation}\label{BianchiF}
\td\Theta^I = 9g C^{IJK} \bar{X}_J \td (f^{-1} X_K) \wedge J, 
\end{equation}  
and
\begin{equation}\label{Maxwell}
\begin{aligned}
\td \star_4 \td (f^{-1} X_I) =& -\frac{1}{6}C_{IJK}\Theta^I \wedge \Theta^J + 2g\bar{X}_I f^{-1} G^- \wedge J  \\ 
&+ 6 g^2 f^{-2}(Q_{IJ} C^{JMN}\bar{X}_M \bar{X}_N + \bar{X}_I X^J \bar{X}_J) \mathrm{dvol}(h).
\end{aligned}
\end{equation}  
For convenience, we also record here an alternate form of the symmetric space condition \eqref{symmcond} as follows
\begin{equation}
\begin{aligned}
C_{IJK}( C^{JLM} C^{KPQ} + C^{JLP} C^{KMQ} + C^{JLQ} C^{KMP}) =& \delta_{IL} C_{MPQ} + \delta_{IM} C_{PQL}   \\ 
&+ \delta_{IP}C_{QLM} + \delta_{IQ} C_{LMP}.
\end{aligned}
\end{equation}

\subsection{The local solution} 
We now present the local form of our solution $(g, F^I, A^I)$.  Our construction has based on the analysis of the supersymmetric solitons of the minimal theory written in terms of the above K\"ahler decomposition \cite{Cassani:2015upa} (this corresponds to setting the $X^I$ constant and setting all the $F^I$ equal) and a similar analysis of the 3-charge solutions discussed in~\cite{Cvetic:2005zi}.

Our starting point is a selection of a K\"ahler base~\cite{Cassani:2015upa}, which we take to be the following orthotoric K\"ahler base metric, which in terms of local coordinates $(y,x,\Phi, \Psi)$ is
\begin{equation}\label{Kahlerbase}
h = g^{-2} \left[ \frac{y-x}{F(x)} \td x^2 + \frac{F(x)}{y-x} (\td\Phi + y \td\Psi)^2 + \frac{y-x}{G(y)} \td y^2 + \frac{G(y)}{y-x}(d\Phi + x \td\Psi)^2\right]
\end{equation} 
for yet-to-be determined single-variable $C^2$-functions $F= F(x)$ and $G= G(y)$. We will work in an open set where $y > x$.  The base metric \eqref{Kahlerbase} actually has a curvature singularity at $y =x$.  We will show in Section 2.3 that the full spacetime metric extends  smoothly across the surface $y=x$ (recall that it is $f^{-1} h$ that appears in the spacetime metric \eqref{BPSmet}). 

The vector fields $\partial_\Phi, \partial_\Psi$ are Killing vector fields of the K\"ahler space, and we will assume they extend to the whole spacetime. The K\"ahler form is
\begin{equation} \label{KahlerJ}
J = g^{-2} \td \left[ (y + x) \td\Phi + x y \td\Psi \right].
\end{equation} 
It is explicitly 
\begin{equation}
g^2 J = \td y \wedge \td\Phi + \td x \wedge \td\Phi + y \td x \wedge \td\Psi + x \td y \wedge \td\Psi .
\end{equation} 
A natural orthonormal frame is
\begin{align}
e^1 &= g^{-1} \left[\frac{y-x}{F(x)}\right]^{1/2} \td x, \quad e^2 = g^{-1} \left[\frac{F(x)}{y-x}\right]^{1/2} (\td\Phi + y \td\Psi), \\ e^3 &= g^{-1} \left[\frac{y-x}{G(y)}\right]^{1/2} \td y, \quad e^4 = g^{-1} \left[\frac{G(y)}{y-x}\right]^{1/2} (\td\Phi + x \td\Psi).
\end{align}  
The orientation is chosen so that $\epsilon_{1234} = -1$, so the volume form $\td \mathrm{vol}_h = -\tfrac{1}{2} J \wedge J$, where 
\begin{equation}
J = e^1 \wedge e^2 + e^3 \wedge e^4.
\end{equation} 
Then this form is obviously anti-self dual.  We will use the same symbol for the $(1,1)$ tensor field $J^a_{~b}$. It can be easily checked that $J^a_{~b} J^b_{~c} = - \delta^a_{~c}$.  The Ricci scalar is easily calculated to be
\begin{equation}
R_h = -\frac{g^2 (f''(x) + g''(y))}{y - x}.
\end{equation} 
We may also identify the other 2 anti self-dual forms:
\begin{equation}
J^2 = e^1 \wedge e^3 - e^2 \wedge e^4 \qquad \text{and} \qquad J^3 = -e^1 \wedge e^4 - e^2 \wedge e^3.
\end{equation} These satisfy the algebra (setting $J^1:=J$)
\begin{equation}
J^i \cdot J^j = -\delta^{ij} \mathbb{I} + \epsilon^{ijk} J^k
\end{equation} 
where $\cdot$ indicates matrix multiplication, $i,j = 1,2,3$ and $\mathbb{I}$ is the identity matrix.  The case $I, J =1$ has been mentioned above. The Ricci form $\mathcal{R}$, defined by
\begin{equation}
\mathcal{R}_{ab} = \frac{1}{2} R_{abcd} J^{cd}
\end{equation} 
is closed and hence locally exact, i.e., $\mathcal{R} = \td \mathcal{P}$, with
\begin{equation}
\mathcal{P}= -\frac{F'(x) (\td\Phi + y \td\Psi) + G'(y)(\td\Phi + x \td\Psi)}{2(y-x)}.
\end{equation}

Having specified the base metric, we now turn to the rest of the fields which determine the full solution.  Let $\mathfrak{q}_I \in \mathbb{R}$ and suppose $\mathfrak{q}_I > 0$.  The scalar fields $X_I$ are chosen to take the form
\begin{equation}
f^{-1} X_I = \frac{\bar{X}_I y + \mathfrak{q}_I}{y-x}.
\end{equation} 
Using \eqref{constraint} we find
\begin{equation}
f = \frac{y-x}{[P(y)]^{1/3}}, \qquad P(y) = y^3 + \alpha_2 y^2 + \alpha_1 y + \alpha_0
\end{equation} where the $\alpha_i$ are defined by
\begin{equation}\label{alpha_i}
\alpha_0 = \frac{9}{2} C^{IJK} \mathfrak{q}_I \mathfrak{q}_J \mathfrak{q}_K, \quad  \alpha_1 = \frac{27}{2} C^{IJK} \bar{X}_I \mathfrak{q}_J \mathfrak{q}_K, \quad \alpha_2 
= \frac{27}{2} C^{IJK} \bar{X}_I \bar{X}_J   \mathfrak{q}_K.
\end{equation} 
The K\"ahler metric $h$ is fully fixed by the choice
\begin{equation}
F(x) = 4\alpha_0 (1-x^2), \qquad G(y) = 4 y (y^2 +  (\alpha_0 + \alpha_2)y + \alpha_1). 
\end{equation} 
Note that he Ricci scalar of $h$ is 
\begin{equation}
R_h = -\frac{8g^2(3y + \alpha_2)}{y-x}, 
\end{equation} 
and it is easily verified that the BPS constraint \eqref{BPSf} is satisfied.  Note the scalars $X^I$ are determined as
\begin{equation}
X^I = \frac{9}{2} C^{IJK} X_J X_K = \frac{\bar{X}^I y^2 + 9 C^{IJK} \bar{X}_J \mathfrak{q}_K y +  \frac{9}{2} C^{IJK}\mathfrak{q}_J \mathfrak{q}_K  }{[P(y)]^{2/3}}.
\end{equation} 
Thus to specify a supersymmetric solution $(g, F^I, X^I)$, it remains to specify the functions $\omega$ and $\Theta^I$, and then actually check that the remaining supersymmetry conditions and field equations are satisfied.  For the self-dual two-forms $\Theta^I$ we take
\begin{equation}
\Theta^I = \frac{ \left(9 C^{IJK} \mathfrak{q}_J \mathfrak{q}_K + (x+y) 9 C^{IJK} \bar{X}_J \mathfrak{q}_K + 2xy \bar{X}^I\right)}{g(y-x)^2}\left[ \td y \wedge (\td\Phi + x \td\Psi) - \td x \wedge (\td\Phi + y \td\Psi) \right].
\end{equation} 
It is evident that these forms are self-dual by writing them in terms of the orthonormal frame $\{e^a\}$.  A long but straightforward computation verifies that the necessary conditions \eqref{ricciform} and \eqref{BianchiF} are automatically satisfied. To perform this calculation, note that 
\begin{equation}
\td \left[ (y-x)^{-2} \td y \wedge (\td\Phi + x \td\Psi) - \td x \wedge (\td\Phi + y \td\Psi) \right] =0.
\end{equation} 
For the one-form $\omega$, we take
\begin{equation}\label{omega}
\omega = \omega_\Phi d\Phi + \omega_\Psi d\Psi, 
\end{equation} 
where 
\begin{align}
\omega_\Phi &= \frac{2}{g(y-x)^2} \left[ \alpha_0 (1-x^2 + x y) + \alpha_1 y + y^2 ( y + \alpha_2) \right] ,\\
\omega_\Psi &= \frac{2y}{g(y-x)^2} \left[ \alpha_0 + \alpha_1 x + \alpha_2 x y + x y^2 \right].
\end{align} 
Recall that supersymmetry requires $X_I \Theta^I = -\frac{2}{3} G^+$. This requires calculating $\star \td\omega$. It is useful to record that the volume form in the $(y,x,\Phi, \Psi)$ coordinates is given by
\begin{equation}
\text{dVol}(h) = \frac{y-x}{g^4}  \td y \wedge \td x \wedge \td\Phi \wedge \td\Psi.
\end{equation} 
and the inverse metric components are
\begin{gather}
h^{yy} = \frac{g^2 G(y)}{y-x}, \qquad h^{xx} = \frac{g^2 F(x)}{y-x}, \qquad h^{\Phi\Phi} = \frac{g^2 (G(y) x^2 + F(x) y^2)}{F(x) G(y) (y-x)}, \\
h^{\Phi\Psi} = - \frac{g^2 (G(y) x + F(x) y)}{F(x) G(y)(y-x)}, \qquad h^{\Psi\Psi} = \frac{g^2 (F(x) + G(y))}{F(x) G(y)(y-x)},
\end{gather} 
and hence we compute that
\begin{align}
\star (\td x \wedge \td\Phi) &= \frac{ G(y) x^2 + F(x) y^2}{G(x)(y-x)} \td y \wedge \td\Psi + \frac{G(y) x + F(x) y}{G(y)(y-x)} \td y \wedge \td\Phi,  \\
\star (\td y \wedge \td\Phi) &= -\frac{G(y) x + F(x) y}{F(x)(y-x)} \td x \wedge \td\Phi - \frac{G(y) x^2 + F(x) y^2}{F(x) (y-x)} \td x \wedge \td\Psi, \\
\star (\td x \wedge \td\Psi) &= -\frac{G(y) x + F(x) y}{G(y)(y-x)} \td y \wedge \td\Psi - \frac{F(x) + G(y)}{G(x)(y-x)} \td y \wedge \td\Phi, \\
\star (\td y \wedge \td\Psi) &= \frac{F(x) + G(y)}{F(x)(y-x)} \td x \wedge \td\Phi + \frac{G(y) x + F(x) y}{F(x) (y-x)} \td x \wedge \td\Psi.
\end{align} 
We now wish to verify that
\begin{equation}\label{cond2}
f^{-1} X_I \Theta^I = -\frac{1}{3}  (d\omega + \star d\omega),
\end{equation} 
The left-hand side is given by the self-dual two-form
\begin{align}
f^{-1} X_I \Theta^I =& \frac{2}{3g(y-x)^3} \left[(\alpha_1 + (x+y) \alpha_2 + 3 x y)y  + 3\alpha_0 +(x+y) \alpha_1 + x y \alpha_2 \right] \\ \nonumber
& \cdot \left[ \td y \wedge (\td\Phi + x \td\Psi) - \td x \wedge (\td\Phi + y \td\Psi) \right].
\end{align} 
We have once again verified that \eqref{cond2} is satisfied.  It only remains to verify that the Maxwell equation \eqref{Maxwell} holds without any further constraints. This is possible after a tedious calculation, with the use of the identities 
\begin{align}
C_{IJK} \bar{X}^J C^{KPQ} \bar{X}_P q_Q =& \frac{9}{2} C_{IJK} C^{JLM} \bar{X}_L \bar{X}_M C^{KPQ} \bar{X}_P q_Q = \frac{q_I}{3} + \frac{\alpha_2}{3} \bar{X}_I, \\
C_{IJK} C^{JLM} C^{KPQ} \bar{X}_L q_M \bar{X}_P q_Q  =& -\frac{1}{9} C_{IJK} \bar{X}^J C^{KMQ} q_M q_Q + \frac{2}{27} \alpha_1 \bar{X}_I + \frac{2}{27} \alpha_2 q_I, \\
C_{IJK}C^{JLM} C^{KPQ} q_L q_M \bar{X}_P q_Q =& \frac{2}{27} (\alpha_0 \bar{X}_I + \alpha_1 q_I).
\end{align}  
Therefore we have satisfied all the necessary and sufficient conditions to produce a local BPS solution where in particular, the metric takes the canonical form \eqref{BPSmet}. 

\subsection{Global analysis and conserved charges} 
The soliton solutions constructed above have an $SU(2) \times U(1)$ isometry, although this is incompatible with a supersymmetric decomposition~\cite{Cassani:2015upa}. Nonetheless, having the explicit bosonic solution, we can express the metric in a coordinate chart where the symmetry is manifest.  To this end define a new chart $(T, r, \psi, \theta, \phi)$ by
\begin{equation}
r = \left[\frac{y}{g^2 \alpha_0}\right]^{1/2}, \quad \theta = \arccos{x}, \quad T = t, \quad \psi = 4\alpha_0 \Psi, \quad \phi = 4\alpha_0 \Phi - 2g t,
\end{equation} 
so that in particular we have a Killing vector field
\begin{equation}
\frac{\partial}{\partial T} = \frac{\partial}{\partial t} + \frac{g}{2 \alpha_0} \frac{\partial}{\partial \Phi} . 
\end{equation} 
In this coordinate chart, the metric takes the manifestly $SU(2) \times U(1)$-invariant form
\begin{equation}\label{met2}
\begin{aligned}
\td s^2 &= -\frac{H(r)^{1/3} W(r) \td T^2}{\alpha_0 B(r)} + \frac{H(r)^{1/3} \alpha_0 \td r^2}{W(r)} + \frac{r^2 B(r)}{4 H(r)^{2/3}} \left(\td \psi + \cos\theta \td \phi + \Omega(r) \td T\right)^2 \\
& + \frac{H(r)^{1/3}}{4 g^2 \alpha_0} \left( \td \theta^2 + \sin^2\theta \td \phi^2 \right),
\end{aligned}
\end{equation} 
where we have defined
\begin{equation}\label{func}
\begin{aligned}
W(r) &= g^4 \alpha_0^2 r^4 + g^2 \alpha_0 r^2 (\alpha_2 + \alpha_0) + \alpha_1, \qquad B(r) = g^4 \alpha_0^2 r^4 + g^2 r^2 \alpha_2 \alpha_0 + \alpha_1 , \\
H(r) & = \alpha_0 ( g^6 \alpha_0^2 r^6 + g^4 \alpha_2 \alpha_0 r^4 + g^2 \alpha_1 r^2 + 1), \qquad \Omega(r)  = -\frac{2 g \alpha_0}{B(r)} . 
\end{aligned}
\end{equation}  
In these coordinates, we may identify the geometry in the asymptotic region $r \to \infty$ with AdS$_5$ with radius $g^{-1}$ .  This is realized by shifting the radial coordinate as $r^2 = R^2 - \alpha_2/(3 \alpha_0 g^2)$ so that as $R \to \infty$, 
\begin{equation}
\begin{aligned}
\td s^2 & \to -(1 + g^2 R^2 + O(R^{-2})) \td t^2 + \frac{\td r^2}{1 + g^2 R^2 + O(R^{-2})}  \\ & +  \frac{R^2 + O(R^{-2}) }{4} \left(\td\psi + \cos\theta \td\phi + O(R^{-4}) \td T \right ) ^2 + \frac{R^2 + O(R^{-2})}{4} \left( \td \theta^2 + \sin^2\theta \td \phi^2\right), 
\end{aligned}
\end{equation} 
and hence to recover an asymptotically globally AdS$_5$ spacetime we must take $\psi \in (0,4\pi)$, $\phi \sim \phi + 2\pi$, and $\theta \in (0,\pi)$ (with standard coordinate singularities at $\theta = 0, \pi$ corresponding to the poles of the $S^3$).  Note that
\begin{equation}
\left(\frac{\partial}{\partial T}\right)^2 = g_{TT} = - \frac{\alpha_0^2 g^6 r^6 + \alpha_0 g^4(\alpha_0 + \alpha_2) r^4 + \alpha_1 g^2 r^2 + 1} {H^{2/3}} < 0,
\end{equation} 
and hence $\partial_T$ is strictly timelike everywhere.  A similar computation shows that  $(\td T)^2 = g^{TT} < 0$ everywhere, and hence the function $T$ may be identified a time function on the spacetime, which is stably causal.  Timelike surfaces of constant $r > 0$ have topology $\mathbb{R} \times S^3$. 

The functions $W(r), H(r), B(r) > 0$ from the requirement $\alpha_i > 0$, $i =0,1,2$.  From \eqref{alpha_i} it is sufficient to assume $\mathfrak{q}_I >0$ and $C_{IJK} \geq 0$ with at least one positive.   Hence the metric is non-degenerate for all $r > 0$. As $r \to 0$, $g_{\psi\psi} = O(r^2)$ and the Killing vector field $\partial/ \partial \psi$ degenerates. To ensure a smooth degeneration of the $S^1$ generated by this vector field, we must impose the regularity constraint
\begin{equation} \label{reg}
\alpha_0 = \alpha_1 . 
\end{equation} 
This places one algebraic constraint on $\mathfrak{q}_I$.  Thus, a  member of this family of globally smooth asymptotically AdS$_5$ gravitational soliton is parameterized by the $N$ positive real constants $\mathfrak{q}_I$ subject to  \eqref{reg}.  The 2-cycle at $r=0$ is a round $S^2$ of radius $r_{S^2} = (2 g \alpha_0^{1/3})^{-1}$ . With the condition \eqref{reg} the full soliton spacetime metric extends to a global metric on $\mathbb{R} \times \mathbb{CP}^2\setminus \{\mathrm{pt}\}$ where the first factor corresponds to the time direction and the second to spacelike Cauchy surfaces $\Sigma_t$ induced on the level sets of constant $t$.  The topology of $\Sigma_t$ is easiest to read off by noting that the induced metric is toric and the resulting toric diagram (rod structure) is that of $\mathbb{CP}^2$ with one vertex removed corresponding to the point `at infinity'.  Equivalently $\Sigma_t$ has the topology of Taub-Bolt space $O(-1) \to S^2$ , i.e., the tautological bundle over $\mathbb{CP}^1$~\cite{Cassani:2015upa}.

Physically, the 2-cycle is prevented from collapse by the magnetic fluxes
\begin{equation}
\mathcal{D}^I = \frac{1}{2\pi} \int_{S^2} F^I = \frac{9}{2 g \alpha_0} C^{IJK}\mathfrak{q}_J \mathfrak{q}_K.
\end{equation}

The total mass of the spacetime may be computed using the conformal mass of  Ashtekar-Magnon-Das \cite{AMD}.  Setting $\Omega = 1/(g R)$ and defining the conformal metric $\bar{g}_{ab} = \Omega^2 g_{ab}$ with conformal boundary at $\Omega =0$ ( $R \to \infty$) , we define the electric part of the Weyl tensor to be
\begin{equation}
\bar{\mathcal{E}}^a_{b} = (g \Omega)^{-2} \bar{g}^{cd} \bar{g}^{ef} n_d n_f C^a_{~c b e}, 
\end{equation}  
where $n = \td \Omega$. The conserved quantity associated with a Killing vector field $\xi$ is 
\begin{equation}
Q[\xi] = \frac{1}{16\pi g} \int_{S^3} \bar{\mathcal{E}}^a_{~b} \xi^b \td S_a,
\end{equation} 
where the integral is taken over the $S^3$ of radius $g^{-1}$  at conformal infinity with unit timelike normal $\td T$.  The mass is associated to $\xi = \partial_T$, which is non-rotating at infinity. A computation reveals that as $R \to \infty$, the relevant components of the Weyl tensor decay as
\begin{equation}
C^T_{~RTR} = \frac{2 \alpha_2}{g^4 \alpha_0 R^6} + O(R^{-8}),
\end{equation} 
and we obtain the mass
\begin{equation}\label{mass}
E := Q[\partial_T] =  \frac{\pi \alpha_2}{4 g^2 \alpha_0}.
\end{equation} 
Next, consider the electric charges $Q_I$. Note that
\begin{equation}
\star F = -f^{-2} \star_4 \td (X^I f) + e^0 \wedge \left( X^I f \star_4 \td \omega +  \Theta^I + 9f^{-1} g C^{IJK} \bar{X}_J X_K J \right)
\end{equation} 
where $e^0 = f ( \td t + \omega)$. We define
\begin{equation}
Q_I = \frac{1}{8\pi} \int_{S^3_\infty} Q_{IJ} \star F^J, 
\end{equation} 
where the integral is taken over the boundary sphere as $r \to \infty$ on a spatial hypersurface defined by $t = T = $ constant.  One finds as $y \to \infty$ that, pulled back to a surface $y=$ constant, 
\begin{equation}
\star F^J  = \left[ \frac{4\alpha_0}{g^2} (\alpha_2 \bar{X}^I - 9 C^{IJK}  \bar{X}_J \mathfrak{q}_K) + O(y^{-1}) \right] \; \td x \wedge \td \Phi \wedge \td \Psi . 
\end{equation} 
Using the fact that $Q_{IJ}= \tfrac{9}{2} \bar{X}_I \bar{X}_J - \tfrac{1}{2} C_{IJK}\bar{X}^K + O(y^{-1})$ as $y \to \infty$, we find
\begin{equation}
Q_I = -\frac{3 \pi \mathfrak{q}_I}{4 g^2 \alpha_0}.
\end{equation} 
Note that the mass \eqref{mass} of these supersymmetric solutions satisfies the BPS relation 
\begin{equation}
E = |\bar{X}^I Q_I| = \frac{\pi \alpha_2}{4 g^2 \alpha_0}.
\end{equation} 
The angular momentum associated to the Killing vector field $\eta = 2 \partial_\psi$, which has $2\pi-$periodic closed orbits,  is computed from the Komar integral 
\begin{equation}
J = \frac{1}{16\pi} \int_{S^3_\infty} \star \td \eta = \frac{\pi}{2 \alpha_0 g^3}.
\end{equation} 
where we have used the expansion
\begin{equation}
\star \td \eta |_{S^3_\infty} = \left(\frac{1}{2\alpha_0 g^3} + O(r^{-1}) \right) \sin\theta \td \theta \wedge \td \psi \wedge \td \phi . 
\end{equation} 
This corresponds to equal angular momenta in two orthogonal planes of rotation at spatial infinity. Note that the angular momentum associated to the Killing vector field $\partial_\phi$ vanishes. 

\subsection{The $U(1)^3$ supergravity theory} 
Of particular interest in the class of supergravity theories is the $\mathcal{N}=1$ gauged supergravity coupled to two Abelian vector multiplets, which has gauge group $U(1)^3$ that arises as a reduction of Type IIB supergravity on $S^5$ (one keeps the maximal Abelian subgroup $U(1)^3$ of the of maximal 5d $SO(6)$-gauged supergravity).  At least locally, a five-dimensional solution to this theory can be oxidized to a solution of type IIB supergravity theory reduced appropriately on $S^5$. In this special case, the solutions presented here were previously obtained~\cite{Cvetic:2005zi} by performing a BPS limit of a more general family of local supergravity solutions~\cite{Chong:2005hr}.  The theory is recovered by setting $I = i = 1,2,3$ with $\bar{X}_i = 1/3$ (or equivalently $\bar{X}^i = 1$) ,  and $C^{ijk} = |\epsilon_{ijk}|$ where $\epsilon_{123} = \pm 1$ is  totally antisymmetric.  For simplicity, we rescale our dimensionless charge parameters $\mathfrak{q}_i \to \mathfrak{q}_i / 3$.  We then have the simplified expressions
\begin{equation}\label{STUalpha}
\alpha_0 = \mathfrak{q}_1 \mathfrak{q}_2 \mathfrak{q}_3, \qquad \alpha_1 = \mathfrak{q}_1 \mathfrak{q}_2 + \mathfrak{q}_1 \mathfrak{q}_3 + \mathfrak{q}_2 \mathfrak{q}_3, \qquad \alpha_2 = \mathfrak{q}_1 + \mathfrak{q}_2  + \mathfrak{q}_3. 
\end{equation}  
The parameters $\mathfrak{q}_i$ are subject to the regularity condition is simply $ \mathfrak{q}_1 \mathfrak{q}_2 \mathfrak{q}_3 =\mathfrak{q}_1 \mathfrak{q}_2 + \mathfrak{q}_1 \mathfrak{q}_3 + \mathfrak{q}_2 \mathfrak{q}_3$.  The solution then takes the canonical supersymmetric form \eqref{BPSmet}  in the $(t,y,x,\Phi,\Psi)$ coordinates with 
\begin{equation}
f = \frac{y-x}{(H_1 H_2 H_3)^{1/3}}  \qquad \text{and} \qquad H_i = y + \mathfrak{q}_i,
\end{equation} 
and the scalar fields are given by
\begin{equation}
X_i = \frac{f}{3} \cdot \frac{y + \mathfrak{q}_i}{y-x},
\end{equation} 
or equivalently
\begin{equation}
X^1 = \frac{y^2 + (\mathfrak{q}_2 + \mathfrak{q}_3)y  + \mathfrak{q}_2 \mathfrak{q}_3}{(H_1 H_2 H_3)^{2/3}}
\end{equation} with similar expressions for $X^2, X^3$  with the natural permutations of the $\mathfrak{q}_i$.  The K\"ahler base space metric $h$ is given by  \eqref{Kahlerbase} where
\begin{equation}
\begin{aligned}
F(x) &= 4 \mathfrak{q}_1\mathfrak{q}_2 \mathfrak{q}_3 (1 - x^2), \\
G(y) & = 4 y (y^2 + (\mathfrak{q}_1\mathfrak{q}_2\mathfrak{q}_3 + \mathfrak{q}_1 + \mathfrak{q}_2 \mathfrak{q}_3) y + \mathfrak{q}_1  \mathfrak{q}_2 + \mathfrak{q}_1\mathfrak{q}_3 + \mathfrak{q}_2\mathfrak{q}_3.
\end{aligned}
\end{equation}  
The one-form $\omega$ is given by \eqref{omega} where the constants $\alpha_i$ are given by \eqref{STUalpha}.  The Maxwell fields $F^i$ are then determined by \eqref{BPSMaxwell} where the K\"ahler form $J$ is given by \eqref{KahlerJ}, and the self-dual forms $\Theta^i$ by
\begin{equation}
\Theta^1 = \frac{ 2 x y + (x + y) (\mathfrak{q}_2 + \mathfrak{q}_3) + 2 \mathfrak{q}_2 \mathfrak{q}_3}{g (y-x)^2} \left[ \td y \wedge (\td \Phi + y \td \Psi) - \td x \wedge (\td \Phi + y \td \Psi) \right] , 
\end{equation} 
with similar expressions for $\Theta^2, \Theta^3$ with the obvious permutations of the $\mathfrak{q}_i$. 

In the coordinate system $(T, r, \theta, \psi,\phi)$ the metric takes the form \eqref{met2} where the functions $W(r), B(r)$ and $\Omega(r)$ are given by \eqref{func} with the constants $\alpha_i$ given by\eqref{STUalpha} and the function $H$ factors as
\begin{equation}
H(r) =\mathfrak{q}_1 \mathfrak{q}_2 \mathfrak{q}_3 (1 + g^2 \mathfrak{q}_1 \mathfrak{q}_2 r^2)(1 + g^2 \mathfrak{q}_1 \mathfrak{q}_3 r^2) (1 + g^2 \mathfrak{q}_2 \mathfrak{q}_3 r^2).
\end{equation} 
The conserved charges and angular momentum are given by 
\begin{equation}
E =\frac{\pi}{4 g^2 \mathfrak{q}_1\mathfrak{q}_2 \mathfrak{q}_3}(\mathfrak{q}_1 + \mathfrak{q}_2 + \mathfrak{q}_3), \qquad Q_i = - \frac{\pi \mathfrak{q}_i}{4 g^2 \mathfrak{q}_1 \mathfrak{q}_2 \mathfrak{q}_3}, \qquad J = \frac{\pi}{2g^3 \mathfrak{q}_1 \mathfrak{q}_2 \mathfrak{q}_3 },
\end{equation}  
and the angular momentum is associated to the Killing vector field $m = 2 \partial_\psi$. The angular momentum associated to the Killing field $\partial_\phi$ vanishes, $J_\phi =0$.  We note that the parameters $q_i$ used in \cite{Cvetic:2005zi} (c.f. (3.28) of that work) are related to the ones used here by $q_i = \mathfrak{q}_i / (g^2 \mathfrak{q}_i \mathfrak{q}_2 \mathfrak{q}_3)$.  The magnetic dipole fluxes out of the $S^2$ are given by
\begin{equation}\label{dipoles}
\mathcal{D}^i = \frac{|\epsilon^{ijk}| \mathfrak{q}_j \mathfrak{q}_k}{2 g (\mathfrak{q}_i \mathfrak{q}_2 \mathfrak{q}_3)} = \frac{1}{ g \mathfrak{q}_i}.
\end{equation} 
From this and the regularity condition it follows that 
\begin{equation}\label{globalconnection}
\frac{g}{4\pi} \int_{S^2} \sum_{i}^3 F^i  = \frac{g}{2} \sum_{i=1}^3 \mathcal{D}^i = \frac{1}{2}, 
\end{equation} 
which is the Dirac quantisation condition appropriate for a manifold with a spin$^{\mathcal{C}}$ structure. 

Above, we mentioned that the $U(1)^3$ solutions could be locally uplifted to Type IIB supergravity along a Sasaki-Einstein five-manifold $Y_5$.  However, in the case that the five-dimensional spacetime has a non-trivial topology, there will generically be global obstructions to producing a smooth ten-dimensional metric.  Indeed, it was observed in~\cite{Cvetic:2005zi} that the solutions constructed here could not be lifted along the simplest Sasaki-Einstein manifold,  $S^5$ (in contrast, to the three-charge supersymmetric AdS$_5$ black holes \cite{Kunduri:2006ek}). In the equal charge case ($\mathfrak{q}_i = \mathfrak{q}$), the solutions above become solutions of minimal gauged supergravity. It was proved there that in this case, globally regular oxidization was indeed possible along more general \emph{regular} Sasaki-Einstein manifolds~\cite{Cassani:2015upa}. In particular, suppose that $Y_5$  is a circle bundle over a Fano K\"ahler-Einstein $M_4$ with Fano index $I$ (e.g. $\mathbb{CP}^2$ has $I=3$).  Suppose for generality, we allow the coordinate $\psi$ to have period $4\pi / p$ ($p=1$ corresponds to the asymptotically globally AdS$_5$ case we have hitherto assumed).  A regular oxidization can be achieved provided $k p / I \in \mathbb{Z}$ where $k \in \mathbb{Z}$ divides $I$, with $k=1$ if and only if $Y_5$ is simply connected \cite{Cassani:2015upa}.  For a concrete example , take the asymptotically globally AdS$_5$ case $p=1$ and $M_4 = \mathbb{CP}^2$. In this case $k = 3$ and $Y = S^5 / \mathbb{Z}_3$,  with the boundary CFT being a quiver gauge theory living on $\mathbb{R} \times S^3$.  More generally a del Pezzo surface $M_4 = d P_i$, $3 \leq i \leq 9$ has $I = 1$ (hence $k=1$) and the boundary CFT is placed on $S^3 / \mathbb{Z}_p$ for any $p \geq 1$.  

In the general case of unequal charge parameters $\mathfrak{q}_i$, however, we are not aware of a compactification of Type IIB supergravity on a general Sasaki-Einstein manifold $Y_5$ which reduces to the $U(1)^3$ gauged supergravity theory.  A compactification on  $S^5$  is known,  and since the isometry group has a maximal torus of rank 3,  the three gauge fields to be naturally incorporated into the ten-dimensional metric~\cite{Cvetic:1999xp}.  In particular, one takes as Type IIB fields 
\begin{equation}
\begin{aligned}
g_{10} & = W^{1/2} \td s^2_5 + W^{-1/2} \sum_{i=1}^3 (X^i)^{-1} \left[ \td \mu_i^2 + \mu_i^2 (\td \phi^i + g A^i)^2\right], \\
F_5 & = (1 + \star_{10}) \left( 2 g \sum_{i=1}^3 ((X^i)^2 \mu_i^2 - W X^i)\td \text{Vol}_5 - \frac{1}{2g} \sum_{i=1}^3(X^i)^{-1} \star_5 \td X^i \wedge \td \mu_i^2 \right. \\
& \left. \phantom{(1 + \star_{10}) + 2g}+ \frac{1}{2 g^2} \sum_{i=1}^3 (X^i)^{-2} \td \mu_i^2 \wedge (\td \phi^i + g A^i) \wedge \star_5 F^i \right)
\end{aligned}
\end{equation} 
where $W:=\sum_{i}^3 \mu_i^2 X^i > 0$, and $(\mu_i, \phi^i)$ are coordinates on $S^5$ where the `direction cosines' satisfy the constraint $\mu_1^2 + \mu_2^2 + \mu_3^2 =1$.  To cover $S^5$, the angles $\phi^i$ must each be identified with period $2\pi$. More precisely, the 3-torus parameterized by $\phi^i$ is defined by the identifications $T_1: (\phi^1, \phi^2, \phi^3) \sim (\phi^1 + 2\pi, \phi^2, \phi^3)$ with similar expressions for $T_2, T_3$.  Since the compactification is purely local, we can also use\footnote{We thank J. Lucietti for this observation.} this embedding with $S^5 / \mathbb{Z}_p$ replacing $S^5$, provided we define the lattice with the identifications $\hat{T}: (\phi^1, \phi^2, \phi^3) \sim (\phi^1 + 2\pi /p, \phi^2 + 2\pi/p, \phi^3 + 2\pi/p)$ along with any two of the original identifications $T_1, T_2, T_3$. These identifications can be straightforwardly derived by relating the above standard coordinates on $S^5$ with those used in writing $S^5$ as a $U(1)$ bundle over $\mathbb{CP}^2$.  The  Killing vector field $\partial_\psi = \frac{1}{3}(\partial_{\phi^1} + \partial_{\phi^2} + \partial_{\phi^3})$ is non-vanishing and generates the $U(1)$ fibre. The $\mathbb{Z}_p$ quotient corresponds to identifying $\psi \sim \psi + 6\pi/p$, with $S^5$ corresponding to $p=1$.   The ten-dimensional metric extends globally to a smooth manifold provided that the connection on the $\mathbb{T}^3$-bundle is globally defined. This requires 
\begin{equation}
\frac{g}{2\pi}\int_{S^2} F^i =  \frac{k^i}{p}, \qquad k^i \in \mathbb{Z}
\end{equation} 
which from \eqref{dipoles} implies $k_i =p /q_i > 0$. However the regularity constraint \eqref{globalconnection} imposes the condition 
\begin{equation} \label{quant} 
k_1 + k_2 + k_3 = p. 
\end{equation} 
It is clear $p=1$ is not allowed.  In the case of equal charges previously investigated in \cite{Cassani:2015upa}, $k_i = 1$ and $p=3$ corresponding to uplifting on $S^3 / \mathbb{Z}_3$.  We have therefore demonstrated that general members of this family of asymptotically globally AdS$_5$ BPS soliton spacetimes can also be uplifted, provided $q_i = p / k_i$ and \eqref{quant} is satisfied.

\section{Evanescent ergosurface and stable trapping}  
We conclude with a discussion on the classical stability of these solutions. As observed above, there is a strictly timelike Killing vector field $\partial_T$, which is static in the asymptotically globally AdS$_5$ region.  This differs from the supersymmetric Killing vector field $V = \partial_t$ associated to the bilinear of a Killing spinor field $\epsilon$ via $V^a \sim \bar\epsilon \gamma^a \epsilon$.  The Killing vector field $V$ is everywhere causal: 
\begin{equation}
|V|^2 = \left(\frac{\partial}{\partial t}\right)^2 = -f^2 = -\frac{(y-x)^2}{P(y)^{2/3}} \leq 0.
\end{equation} 
and $|V|^2$ vanishes if and only if $y =x$. This relation defines a smooth timelike hypersurface $\mathcal{S}$ along which $V$ is null.  Hence $\mathcal{S}$ may be identified as an \emph{evanescent ergosurface}~\cite{Gibbons:2013tqa, Eperon:2016cdd, Keir:2018hnv}.  The geometry of $\mathcal{S}$ is simplest to see in the $(T,r,\theta,\phi,\psi)$ coordinate chart, as $\partial_T$ remains timelike.  The induced metric on the timelike hypersurface $\mathcal{S}$ is 
\begin{align}
\td s^ 2 = H(r)^{1/3} & \left[-\frac{W(r) \td T^2}{\alpha_0 B(r)} + \frac{H(r) \td r^2}{W(r)(1 - g^4 \alpha_0^2 r^4)} + \frac{r^2 B(r)}{4H(r)} \left( \td \psi + g^2 \alpha_0 r^2 \td \phi + \Omega(r) \td T\right)^2\right. \nonumber \\  +  &\left. \frac{(1 - g^4 \alpha_0^2 r^4) \td \phi^2}{4 g^2 \alpha_0} \right] 
\end{align} 
A simple analysis of the fixed points sets of the torus action generated by $(\partial_\psi, \partial_\phi)$ reveals that the constant $T$-surfaces have $S^3-$topology and are equipped with an inhomogeneous metric.  To see, this observe that the metric in the above coordinates is smooth and positive definite when the radial coordinate takes value in $0 < r < ( g\sqrt{ \alpha_0})^{-1}$.  $\mathcal{S}$ intersects the $S^2$ bubble at $r =0$, where $\partial_\psi$ smoothly degenerates, and at $r = (g  \sqrt{\alpha_0})^{-1}$ the Killing vector field $\partial_\phi - \partial_\psi$ smoothly degenerates. 

The supersymmetric Killing vector field $V$ is easily seen to be tangent to affinely parametrized null geodesics  on $\mathcal{S}$: 
\begin{equation}
\nabla_V V\vert_{\mathcal{S}} = -\frac{1}{2} \td (|V|^2)\vert_{\mathcal{S}}  = 0
\end{equation} 
since $V$ has a second-order zero on $\mathcal{S}$.  With respect to an observer at infinity moving along the orbits of $V$, the conserved energy along these geodesics must vanish because $E = -V \cdot V = 0$ on $\mathcal{S}$, that is, $V$ is tangent to zero-energy geodesics on the evanescent ergosurface.   Such geodesics are also \emph{stably trapped}  as proved in~\cite{Eperon:2016cdd}.   We will briefly overview their elegant argument, which employs the Jacobi equation for geodesic deviation and the fact that $V$ is a Killing vector field such that $-|V|^2$ is minimized on $\mathcal{S}$.  

Given a energy-minimizing null geodesic $\gamma$ on $\mathcal{S}$ with tangent vector field $V$, consider a one-parameter family of causal geodesics $\gamma_s$ with $\gamma_0 = \gamma$ and associated causal tangent vector fields $X_s$ and geodesic deviation vector field $Y$. The geodesic deviation equation on $\gamma$
\begin{equation}
\nabla_V \nabla_V Y^a |_\gamma = R^a_{~bcd} V^b V^c Y^d|_\gamma
\end{equation} 
admits a first integral as a consequence of the Killing property of $V$
\begin{equation} \label{firstint}
|\mathcal{L}_V Y|^2 + H_{ab} Y^a Y^b = C,
\end{equation} 
where $C$ is a constant on $\gamma$, and $H_{ab}$ is the Hessian of $-V^2/2$, i.e., $H_{ab} = \nabla_a \nabla_b (-V^2/2)$.  Using the fact that $|X_s|^2$ is maximized on $\gamma$, it can be shown that $V\cdot Y$ is a constant on $\gamma$. Thus 
\begin{equation}
V \cdot \mathcal{L}_V Y =0,
\end{equation} 
from which it follows that $\mathcal{L}_V Y$ is spacelike or null on $\gamma$.  Hence, the first term in \eqref{firstint} is non-negative. Moreover a direct computation shows that
\begin{equation}
H_{ab} = 4g^2 n_a n_b,
\end{equation} 
where the spacelike unit normal to $\mathcal{S}$ is given by
\begin{equation}
n = \frac{1}{2 g H^{1/3}} \td \left( g^2 \alpha_0 r^2 - \cos\theta \right).
\end{equation} 
Hence $H_{ab}$ is a positive-definite metric on the space on the vectors normal to $\mathcal{S}$ and vanishes on those vectors tangent to $\mathcal{S}$.  In particular, \eqref{firstint} implies that $C \geq 0$ and
\begin{equation} %%% fadsfsdafsda
H_{ab} Y^a Y^b = 4g^2 ( n \cdot Y)^2 \leq C,
\end{equation} 
where $ n \cdot Y$ measures the component of the deviation vector field $Y$ normal to $\mathcal{S}$.  Therefore the normal component of $Y$ remains bounded on the evanescent ergosurface. This demonstrates the stable trapping property, namely,  that initially nearby causal geodesics  to $\gamma$ remain sufficiently close. 

The phenomena of stable trapping provide a geometric obstruction to the establishment of sufficiently strong decay statements for solutions of wave equations~\cite{Eperon:2016cdd, Keir:2018hnv, Keir:2014oka}.   Intuitively, since the behavior of high-frequency waves can be approximated by null geodesics, the trapping property should lead to the clumping of energy in a bounded region.  In particular, a fast (e.g., polynomial) decay of such solutions is widely expected to be required for nonlinear stability.  On the other hand, unstable trapping, such as that which occurs at the photon sphere $r = 3M$ of the Schwarzschild spacetime, is known not to prevent sufficiently fast decay.  As shown above for the gravitational solitons considered here, however, the trapping is stable, and in addition, there is no event horizon to aid in decay.

In the stationary, asymptotically flat case, Moschidis has rigorously proved that provided an energy boundedness statement is true for solutions of the linear wave equation, then the local energy of waves must decay at least inverse logarithmically~\cite{Moschidis:2015wya}.  The question arises whether, in a given spacetime,  such a decay statement is sharp or whether one can prove faster decay.  For supersymmetric microstate geometries with evanescent ergosurfaces (and hence fall outside the hypotheses of Moschidis' theorem), Keir rigorously established that a stronger decay statement cannot exist and indeed any spacetime possessing an evanescent ergosurface but no event horizon exhibits a linear instability~\cite{Keir:2018hnv} (see also Keir's analysis of a particular family of supersymmetric microstate geometries \cite{Keir:2016azt}).  Analogous results have been established for non-supersymmetric microstate geometries that exhibit stable trapping despite not having an evanescent ergosurface \cite{Gunasekaran:2020pue} (in fact, the energy of solutions to the linear wave equation are uniformly bounded).

In the asymptotically Anti-de Sitter case, slow decay of waves caused by stable trapping has been investigated by Holzegel and Simulivici in Kerr-AdS$_4$ spacetimes~\cite{Holzegel:2011uu, Holzegel:2013kna}. Here the underlying geometric obstruction to decay is a combination of unstable trapping near the horizon and the lack of dispersion at null infinity (assuming standard reflective boundary conditions).  In the present case for the family of asymptotically globally AdS$_5$ solitons considered there, one has both the combined effect of stable trapping at an evanescent ergosurface along with the usual lack of dispersion at infinity, and moreover, there is no horizon to help with decay. These considerations strongly suggest that these solutions, despite being supersymmetric, are nonlinearly unstable and likely also unstable at the linear level, as is generically the case for similar asymptotically flat and asymptotically Kaluza-Klein spacetimes~\cite{Keir:2018hnv}.  Indeed, even the maximally supersymmetric AdS vacuum is now known to be nonlinearly unstable to arbitrarily small, generic perturbations that lead to the formation of black holes \cite{Bizon:2011gg, Bizon:2015pfa, Dias:2011ss},  and therefore it is reasonable to expect taht the endpoint of instability of supersymmetric globally AdS$_5$ solitons would be a non-BPS asymptotically AdS$_5$ black hole spacetime.   As a first step, it would be useful to prove that general solutions to the Klein Gordon equation on these soliton backgrounds cannot decay faster than logarithmically, following the quasimode construction strategy of~~\cite{Holzegel:2011uu, Holzegel:2013kna}.

\paragraph{Acknowledgements} We thank James Lucietti for several useful suggestions. HKK acknowledges the support of the NSERC Grant RGPIN-2018-04887.

\end{document}